\documentstyle[12pt,epsfig]{article}
\def\lsim{\raise0.3ex\hbox{$<$\kern-0.75em\raise-1.1ex\hbox{$\sim$}}}
\def\gsim{\raise0.3ex\hbox{$>$\kern-0.75em\raise-1.1ex\hbox{$\sim$}}}

\def\beq{\begin{equation}}
\def\eeq{\end{equation}}
\def\bea{\begin{eqnarray}}
\def\eea{\end{eqnarray}}
\def\bq{\begin{quote}}
\def\eq{\end{quote}}

\newcommand{\rr}{\mbox{\boldmath $r$}}

\def\gappeq{\mathrel{\rlap {\raise.5ex\hbox{$>$}}
{\lower.5ex\hbox{$\sim$}}}}

\def\lappeq{\mathrel{\rlap{\raise.5ex\hbox{$<$}}
{\lower.5ex\hbox{$\sim$}}}}

\def\Toprel#1\over#2{\mathrel{\mathop{#2}\limits^{#1}}}

\begin{document}
\pagestyle{empty}
\begin{center}
{\bf INVESTIGATING $F_L(x,Q^2)$  AT FIXED  ENERGY IN THE COLOR DIPOLE FORMALISM}

\vspace*{0.7cm}
 M.V.T. Machado$^{\,1,2}$ \\
\vspace{0.3cm}
{\rm $^{1}$ Universidade Estadual do Rio Grande do Sul (UERGS) \\
Unidade de Bento Gon\c{c}alves, CEP 97500, Bento Gon\c{c}alves, Brazil\\
\rm $^{2}$ High Energy Physics Phenomenology Group, GFPAE,  IF-UFRGS \\
Caixa Postal 15051, CEP 91501-970, Porto Alegre, RS, Brazil}
\vspace*{1cm}

{\bf ABSTRACT}
\end{center}

\vspace*{0.8cm} \noindent


\vspace{-1.cm} \setcounter{page}{1} \pagestyle{plain}

At small-$x$, the structure function  $F_L(x,Q^2)$ is driven by the gluon content of the nucleon target and consequently it can unravel the underlying QCD dynamics in that region. In this work, one studies its behavior on photon virtuality $Q^2$ at fixed energy  within the color dipole formalism for models considering parton saturation effects. The reason is that they resum a wide class of higher-twist contributions, which have important influence in $F_L$ description towards low $Q^2$. It is shown that the geometric scaling property holds for the longitudinal cross section. Moreover, the effective anomalous dimension in scaling dipole cross sections can be investigated by studing both the turn over and the large $Q^2$ regions of the recent experimental measurements.

\noindent \rule[.1in]{17cm}{.005in}

\vspace{0.5cm}

\section{Introduction}

The longitudinal
structure function $F_L$ at high energies can be written in terms of the cross section for the absorption of longitudinally polarized photons.  Its accurate  measurements at low Bjorken variable $x$ and/or photon virtuality $Q^2$ would be helpful to constrain the underlying physics in that kinematical region. Namely, it is expected that it may discriminate between the leading twist
predictions, which consider the collinear factorization and parton
distributions determined from global fits, and the predictions
from the saturation models which resum a class of higher twist
contributions at small $x$. Along these lines, recently it has been claimed \cite{Thorne_new} that an accurate direct measurement could teach us the best way in which to use perturbative QCD for structure functions, since it gives an independent test of the gluon distribution at low $x$. It would be a much direct test of the success of different theories in QCD. The motivation for the following study is the recent extraction of $F_L$ at fixed energy by H1 Collaboration at HERA \cite{h1_prel}.

Lets shortly summarize the present status of the perturbative QCD calculation  using the collinear factorization. The $F_L$ is well understood at high $Q^2$ whereas very little
 is known about its behavior towards  low $Q^2$ and small $x$. Theoretically, we have that in the limit
$Q^2 \rightarrow 0$ the structure function  $F_L$ has to vanish as
$Q^4$ reflecting the fact that the interaction  of
longitudinally polarized virtual photons  has to vanish  in the photoproduction limit. On the other hand, the leading twist DGLAP  MRST \cite{mrst} and CTEQ \cite{cteq} global fits require the gluon distribution  to be valence-like or negative at
small $x$ and low $Q^2$ in order to describe the experimental data, leading to $F_L$ being negative at the smallest $x$-$Q^2$ bins. At that region, a comparison of the predictions  at LO, NLO and NNLO using
MRST partons have shown a poor description of the experimental results \cite{thorne}.  These results demonstrate the limitations of the applicability of perturbation theory and the necessity of resummation procedures. Nevertheless, in
the global fit of the existing light-targets DIS data at LO, NLO
and NNLO QCD approximations performed in Ref. \cite{alekhin}, the high-twist contributions to the structure functions  have been estimated. It was verified that these terms do not vanish up to NNLO and
give important contributions at both small and large $x$ regions.
Therefore, a direct analysis of $F_L$ could  discriminate
between leading twist and higher-twist resummations.

There are also interesting and successful phenomenological analyses using the semihard factorization approach \cite{Kotikov}. The longitudinal structure function is now obtained as a convolution of the structure function of the off-shell gluons with virtuality $k_{\perp}$ with the unintegrated gluon distribution.  The data description is good, with the predictions being intermediate between the LO and NLO collinear calculations. The agreement is even improved if non-linear effects are included in the gluon distribution and/or NLO corrections to the approach are introduced. For the saturation effects, the GLLM unintegrated gluon distribution \cite{GLLM} has been considered, whereas an estimation for high order effects is taken into account by changing the hard scale in the strong coupling constant $\alpha_s(\mu^2)$, where $\mu^2 = cQ^2$ ($c\,\gg 1$). The GLLM results for dipole cross section/unintegrated gluon function \cite{GLLM} are obtained by a global QCD analysis using the numerical solution for the Balitsky-Kovchegov nonlinear equation, which sums higher twists while preserving unitarity, and the short distance contribution is given by the DGLAP kernel. It has been shown that higher-twist contributions in the form of saturation effects result in better data description. This motivates an analysis on the dipole formalism \cite{dipole}, as $k_{\perp}$-factorization approach and the color dipole approach are equivalent at the leading logarithmic approximation. 

An advantage in using color dipole models is that there are several analytical expressions for the dipole cross section, which describes the interaction of color dipoles with the (proton) target. This allows to obtain very fast calculations for the observables of interest and perform important qualitative studies on their behavior. In particular, it has  been observed that the DESY-HERA data at small $x$ and low $Q^2$ can be successfully described with the help of saturation models. Moreover, the total cross section \cite{scaling} and also  the  inclusive charm production \cite{prl} present the property of  geometric scaling, which is a characteristic feature  of the high density QCD approaches.
The saturation (non-linear QCD) approaches are characterized by a typical  scale, denoted the saturation scale $Q^2_{\mathrm{sat}}(x)\propto x^{-\lambda}$, which is energy
dependent and marks the transition between the  linear (leading twist) perturbative QCD regime and saturation domain. The phenomenological models indicate that the saturation scale is smaller than 2 GeV$^2$ at HERA and thus one expects that the signatures of the saturation effects become increasingly  evident in the
region of small $x$ and very low $Q^2$. Furthermore, some of  these approaches contain information of all orders in $1/Q^2$, namely they resum higher twist
contributions \cite{peters,gotsman}. These corrections should be important at the low $Q^2$ region, where the leading twist  approaches would be in the limit of their applicability. These effects to $F_L$ has been recently investigated within the dipole picture using saturation models at Ref. \cite{vicmag_fl}. Early related analysis on longitudinal structure function within the dipole approach can be also found in Refs. \cite{Nikolaev:1994ce,Nikolaev:1994bg} and in Ref. \cite{Ivanov:2000cm} for a recent investigation using semihard approach.

In this Letter, one analyzes the behavior of the longitudinal
structure function $F_L(x,Q^2)$ at fixed energy as a function of $Q^2$. One  considers the color dipole formalism, focusing mainly on the saturation models. In particular, one investigates the analytical solutions for the color dipole cross sections presenting geometric scaling property. We also call  attention that the longitudinal cross section at small-$x$ should exhibit scaling pattern as the inclusive case and this could be demonstrated from available experimental data. This Letter is organized as follows. In the next section, the theoretical description of $F_L$ within the color dipole formalism is reviewed, introducing several implementations for the dipole cross section. In Sec. 3, one computes analytical expressions for $F_L$ using scaling dipole cross section in the regions of large $Q^2$ and in the turn-over region, which are useful in the determination of the effective anomalous dimension. In the last section, one presents the results for $F_L$ at fixed energy and as a function of $Q^2$, which can be contrasted with the recent experimental results. Moreover, one demonstrates that $\sigma_L$ should naturally exhibit geometric scaling property and theoretical estimations are presented. Conclusions and a summary are also presented in last section.

\section{Structure function $F_L$ and color dipole approach}

In the color dipole formalism, the deep inelastic scattering process can be seen as a succession in time of three factorisable subprocesses: i) 
the photon fluctuates in a quark-antiquark pair with transverse separation $\rr \sim 1/Q$ long after the interaction, ii) this 
color dipole interacts with the proton target, iii) the quark pair
annihilates in a virtual photon. The longitudinal structure function is related to the longitudinal $\gamma^* p $ cross section as $F_L= \frac{Q^2}{4\pi^2 \alpha_{em}} \sigma_L^{\gamma^* p}$. The latter is the overlap of the dipole cross section on the longitudinal photon wavefunction. The interaction is then factorized in the simple formulation \cite{dipole},
\begin{eqnarray}
F_L\,(x,Q^2)  =  \frac{Q^2}{4\pi^2 \alpha_{\mathrm{em}}}\,\int dz \,d^2\rr\,
|\Psi_{L}\,(z,\,\rr;\,Q^2)|^2
\,\sigma_{dip}\,(\tilde{x},\,\rr),
\label{FLdip}
\end{eqnarray}
where $z$ is the longitudinal momentum fraction of the quark in the color dipole, $\tilde{x} = \frac{ Q^2+m_q^2}{W_{\gamma p}^2+Q^2}$ is equivalent to the Bjorken
variable and provides an interpolation for the $Q^2\rightarrow 0$ limit. The mass of the quark of flavour $f$ is labeled as $m_f$.  The longitudinal photon wavefunctions $\Psi_{L}$ is determined
from light cone perturbation theory and read as
\begin{eqnarray}
 |\Psi_{L}\,(z,\,\rr;\, Q^2)|^2 =  \frac{6\,\alpha_{\mathrm{em}}}{\pi^2} \,
\sum_f e_f^2 \, \left[Q^2 \,z^2 (1-z)^2
\,K_0^2\,(\varepsilon\,\rr) \right]\,, 
\label{wlongs}
 \end{eqnarray}
where the auxiliary variable $\varepsilon^2=z(1-z)\,Q^2 + m^2_f$ and $K_{0}$ is  the modified Bessel function. Finally, a summation over the quark flavors is understood in Eq. (\ref{FLdip}). It should be noticed that $F_L$ goes to zero when $Q^2 \rightarrow 0$ at low $x$ in the dipole picture since $|\Psi_L|^2 \propto Q^2$. The dipole hadron cross section $\sigma_{dip}$  contains all
   information about the target and the strong interaction physics.
There are several phenomenological implementations for this
quantity and the main feature is
to be able to match the soft (low $Q^2$) and hard (large $Q^2$)
regimes in an unified way. 

In the calculation presented here one considers analytical expressions for the dipole cross section, with particular interest for those ones presenting scaling behavior. Namely, one has $\sigma_{dip} \propto (\rr^2Q^2_{\mathrm{sat}})^\gamma$ for dipole sizes $\rr^2 \approx 1/Q^2_{\mathrm{sat}}$ and where $\gamma$ is the effective anomalous dimension. In what follows one takes the phenomenological parameterizations: (a) Golec-Biernat-W\"{u}sthoff model (GBW) \cite{golecwus}, (b) Itakura-Iancu-Munier (IIM) model \cite{iancu_munier} and (c) Kharzeev-Kovchegov-Tuchin (KKT) \cite{kkt05}. In order to investigate the DGLAP evolution for small dipoles, the Bartels-Golec-Bienat-Kowalski (BGBK) model \cite{bgbk} is also considered. We call attention for the recent GLLM dipole cross section \cite{GLLM}, which contains saturation corrections via solution of Balitsky-Kovchegov equation and DGLAP kernel for the linear regime. It is not included here as we are interested in analytical parameterizations. Let us summarize the main features of each parameterization. The GBW parameterization takes the eikonal-like form,
\begin{eqnarray}
\sigma_{dip} \,(\tilde{x}, \,\rr^2)  =  \sigma_0 \, \left[\, 1- \exp
\left(-\frac{\rr^2Q_{\mathrm{sat}}^2}{4} \right) \, \right]\,,
\hspace{1cm} Q_{\mathrm{sat}}^2\,(x)  =  \left( \frac{x_0}{\tilde{x}}
\right)^{\lambda} \,\,\mathrm{GeV}^2\,. \label{gbwdip}
\end{eqnarray}
 where the parameters were obtained from a fit to the HERA data producing $\sigma_0=23.03 \,(29.12)$ mb, $\lambda= 0.288 \, (0.277)$ and $x_0=3.04 \cdot 10^{-4} \, (0.41 \cdot 10^{-4})$ for a 3-flavor
(4-flavor) analysis~\cite{golecwus}. An
additional parameter is the effective light quark mass, $m_f=0.14$
GeV, which plays the role of a regulator at the photoproduction limit. The charm mass is set to be $m_c=1.5$ GeV. The GBW parameterization presents a geometric scaling form, $\sigma_{dip}\propto f(\rr^2Q_{\mathrm{sat}}^2)$. For small dipoles $\rr^2\leq 1/Q_{\mathrm{sat}}^2$ it can be approximated by $\sigma_{dip} \simeq \sigma_0 (\rr^2Q_{\mathrm{sat}}^2/4)$, where the effective anomalous dimension is equal one, $\gamma =1$. The BGBK model is the implementation of QCD evolution in the the dipole cross section, which  depends on the gluon distribution as, \begin{eqnarray}
 \sigma_{dip}   =  \sigma_0 \, \left[\, 1- \exp
\left(-\frac{\,\pi^2\,\rr^2\,\alpha_s(\mu^2)\,\tilde{x}\,G\,(\tilde{x},\mu^2)}{3\,\sigma_0}
\right) \, \right]\,,\hspace{0.6cm} x\,G(x,\,Q_0^2) = A_g\,x^{-\lambda_g}\,(1-x)^{5.6}
\label{bgkdip} 
\end{eqnarray} 
where  the initial condition is taken at $Q_0^2=1$ GeV$^2$ and the hard scale is $\mu^2=C/\rr^2 + \mu_0^2$. The phenomenological parameters are determined from a fit to small $x$ HERA  data. The function $G(x,\mu^2)$ is evolved with the
leading order DGLAP evolution equation for the gluon density. The improvement preserves the main features of the
low-$Q^2$ and transition regions, while providing QCD evolution in the
large-$Q^2$ domain. The effective anomalous dimension for the BGBK is expected to be close to the DGLAP anomalous dimension, $\gamma_{\mathrm{DGLAP}} = 1$.

Although GBW and BGBK models are very successful in describing HERA data,  their functional forms are only an approximation of the theoretical non-linear QCD approaches. An analytical expression for the dipole cross section can be obtained within the BFKL formalism and intense theoretical studies has been performed towards an understanding of the BFKL approach in the border of the saturation region \cite{IANCUGEO,MUNIERWALLON}. In particular, the dipole cross section has been calculated in both LO  and NLO BFKL  approach in the geometric scaling region \cite{BFKLSCAL}. The IIM parameterization smoothly interpolates between the  limiting behaviors analytically under control: the solution of the BFKL equation
for small dipole sizes, $\rr\ll 1/Q_{\mathrm{sat}}(x)$, and the Levin-Tuchin law \cite{Levin}
for larger ones, $\rr\gg 1/Q_{\mathrm{sat}}(x)$. A fit to the structure function $F_2(x,Q^2)$ was performed in the kinematical range of interest.  The IIM dipole cross section is parameterized as follows,
\begin{eqnarray}
\sigma_{dip}\,(x,\,\rr) = \sigma_0 n_0\left(\frac{\rr^2 Q_{\mathrm{sat}}^2}{4}\right)^{\gamma_{\mathrm{sat}} + \frac{\ln \,(2/r Q_{\mathrm{sat}})}{\kappa \,\lambda \,Y}}\!\Theta\left(\rr - R_{\mathrm{sat}}\right)\, + \,\sigma_0\left[1 - e^{-a\ln^2\,(b\,\rr\, Q_{\mathrm{sat}})}\right]\Theta \left(R_{\mathrm{sat}}-\rr\right)\,,
\label{CGCfit}
\end{eqnarray}
where $R_{\mathrm{sat}}\,(x) = 2/Q_{\mathrm{sat}}$ and the expression for $\rr Q_{\mathrm{sat}}(x)  > 2$  (saturation region)   has the correct functional
form, as obtained either by solving the Balitsky-Kovchegov (BK) equation \cite{bal,kov}, 
or from the theory of the Color Glass Condensate (CGC) \cite{CGC}.  The coefficients $a$ and $b$ are determined from the continuity conditions of the dipole cross section  at $\rr Q_{\mathrm{sat}}=2$. The coefficients $\gamma_{\mathrm{sat}}= 0.63$ (the BFKL anomalous dimension at the saturation border) and $\kappa= 9.9$  are fixed from their LO BFKL values. The parameters used here are $\sigma_0=2\pi R_p^2= 26$ mb, $\lambda=0.253$, $x_0=0.267\times 10^{-4}$ and $n_0=0.7$. The light quark mass is set to be $m_f=0.14$ GeV in similar way as GBW model. The IIM parameterization presents scaling violation once the effective anomalous dimension depends also on the rapidity $Y=\ln(1/x)$ for small size dipoles, $\gamma\,(x,\rr)=\gamma_{\mathrm{sat}} + \frac{\ln (2/\rr Q_{\mathrm{sat}})}{\kappa \,\lambda \,Y}$. For small dipoles having transverse size relatively close to $R_{\mathrm{sat}}$ the second term vanishes and the dipole cross section scales as $\sigma_{dip}\propto \sigma_0\,(\rr^2Q_{\mathrm{sat}}^2/4)^{\gamma_{\mathrm{sat}}}$.

The KKT parameterization has been  proposed  in order to describe  hadron 
production in dAu collisions  at forward and mid-rapidities. The  phenomenological model for the quark and gluon dipole  scattering amplitudes is inspired in the approximated analytical solutions of the BK equation for the saturation and color transparency regimes. The expression for the quark dipole-target forward scattering amplitude is given by \cite{kkt05}:
\begin{eqnarray}
\sigma_{dip}\,(x,\,\rr) = \sigma_0\, \left\{1- \exp\left[-\frac{1}{4} \left(\rr^2
\frac{C_F}{N_c} \, Q_{\mathrm{sat}}^2\right)^{\gamma(x,\,r)}\right]\right\}\,,
\end{eqnarray}
where   the effective  anomalous dimension $\gamma(x,\,\rr)$  is parameterized as,
\begin{eqnarray}
\label{gamma}
\gamma(x, \,\rr) \, = \, \frac{1}{2}\left(1+\frac{\kappa 
(x, \rr)}{\kappa (x,\rr) + \sqrt{2 \,\kappa (x, \rr)}+  7
\zeta(3)\, c} \right),\hspace{0.7cm} \kappa\, (x, \,\rr) \, = \, \frac{\ln\left[1/( \rr^2 \, Q_{s0}^2 ) 
\right]}{(\lambda/2)(Y-Y_0)}\,.
\end{eqnarray}
where $c$ is a free parameter and $Q_{\mathrm{sat}}^2(x)  = \Lambda^2 \,x^{-\lambda}$, with $\Lambda = 0.6$ GeV and $\lambda=0.3$.  The initial saturation scale used in the expression for $\kappa(x,\rr)$ is defined by $Q_{s0}^2=Q_{\mathrm{sat}}^2(Y_0)$ with $Y_0$ being  the
lowest value of rapidity at which the low-$x$ quantum evolution
effects are essential. The form of the anomalous
dimension is inspired by the analytical solutions to the BFKL equation. That is, in the limit $\rr\rightarrow 0$ at fixed $x$ one 
recovers the anomalous dimension in the double logarithmic
approximation $\gamma \approx 1 - \sqrt{1/(2 \, \kappa)}$. In another
limit of large $Y$ with $\rr$ fixed, Eq. (\ref{gamma}) reduces to the
expression of the anomalous dimension near the saddle point in the
leading logarithmic approximation $\gamma \approx
\frac{1}{2} + \frac{\kappa}{14 \, c \, \zeta (3)}$. Therefore, $\gamma (x,\rr)$
mimics the onset of the geometric scaling region. In Ref. \cite{kkt05} the authors assume a characteristic value  $\rr
\approx 1/(2 \, k_{\perp})$, where $k_{\perp}$ is the transverse momentum of the valence quark, and the anomalous dimension  is approximated by 
$\gamma(x, \rr) \approx \gamma(x,1/(4 \, k_{\perp}^2))$. The parameters for KKT dipole cross section has been determined from the inclusive hadron production in deuteron-gold collisions at $\sqrt{s} = 200$ GeV at RHIC ($Y_0=0.6$ and $c=4$). However, it has been shown in Ref. \cite{KGN} that those parameters are not able to describe the DESY-HERA data on structure functions. The main reason is that those parameters lead to a vary fast transition to saturation region and/or the saturation scale is somewhat larger (it goes to unity already at $x=3\cdot 10^{-2}$) than GBW and IIM models.

In order to obtain more reliable results with KKT model, in the analysis presented here the following modifications will take place. The saturation scale is taken from GBW model and the overall normalization is set as $\sigma_0^{\mathrm{KKT}}=(N_c/C_F)\,\sigma_0^{\mathrm{BGW}}$. With this procedure, the KKT model reproduces the GBW parameterization in the region where $\gamma (x,\rr)=1$. Moreover, for computing $\kappa (x,\,\rr)$ one takes the identification $\rr^2=1/Q^2$, $Y_0 = 4.6$ (which corresponds to $x=10^{-2}$) for the lowest value of rapidity at which quantum evolution is essential and $c=0.2$. Putting together these ingredients, the KKT dipole cross section becomes similar to the BGBK dipole cross section.  This fact seems to be consistent as BGBK has an effective anomalous dimension close to the anomalous dimension in the double logarithmic approximation for small size dipoles $\rr \rightarrow 0$.

In next section, the analytic behavior of the longitudinal structure function at fixed energy as a function of $Q^2$ is computed for dipole cross sections presenting scaling property. These results could help on a better understanding of the anomalous dimension. We will look at the large $Q^2 $ region, where small dipole configurations dominate, and also at the intermediate region of virtuality where saturation effects become important.

\section{Analytical results and effective anomalous dimension}

Let us start analyzing the large $Q^2$ region. It is dominated by small size dipoles having transverse sizes lower than $1/Q_{\mathrm{sat}}$. In this region, the dipole cross section in the models presented above takes the approximate form $\sigma_{dip}\propto (\rr^2\,Q_{\mathrm{sat}}^2)^\gamma $. Hence, at the limit $\rr\rightarrow 0$ GBW model gives $\gamma =1 $, the IIM model gives $\gamma \rightarrow 1$ (a rough pure scaling fit gives $\gamma \approx 0.84$ for the range $Q^2\leq 45$ GeV$^2$) and KKT provides $\gamma \approx 1 - \sqrt{1/(2 \, \kappa)}$. Moreover, models based on BFKL evolution equation present an anomalous dimension $\gamma_{\mathrm{BFKL}}=1/2$.  Using the scaling approximation for the dipole cross section it can be shown that,
\begin{eqnarray}
\int\! dz \,d^2\rr \left(\rr^2Q_{\mathrm{sat}}^2\right)^{1-\gamma_s}
|\Psi_{L}|^2 = \frac{4\,\alpha_{\mathrm{em}}N_c}{\pi}\left(\frac{Q_{\mathrm{sat}}^2}{Q^2}\right)^{1-\gamma_s}\!\sum_f e_f^2\frac{(1-\gamma_s)}{4^{\gamma_s}}
\frac{\Gamma^2(1+\gamma_s)\Gamma^2(1-\gamma_s)\Gamma^2(2-\gamma_s)}
{\Gamma(2-2\gamma_s)\Gamma(2+2\gamma_s)(3-2\gamma_s)}\,,
\label{analytic} 
\end{eqnarray}
where $\gamma = 1-\gamma_s$ and the expression above is obtained in the massless limit. 

Inserting the result of expression Eq.(\ref{analytic}) into Eq. (\ref{FLdip}), the large $Q^2$ behavior for the longitudinal structure function can be explicitly computed. For most of models $\gamma=1$, whereas for BFKL approach $\gamma =1/2$. Thus, $\gamma_s=0$ for GBW, BGBK and IIM (and probably for KKT) and $\gamma_s=1/2$ for LO BFKL approaches.  Therefore, one obtains,
\begin{eqnarray}
F_L\,(x,Q^2) \approx \left\{ \begin{array}{ll} 
 \frac{\sigma_0}{4\pi^3}\sum e_f^2\,Q_{\mathrm{sat}}^2\,(x)\,,& \mbox{for $\gamma=1$ (GBW) }\,,\\
\frac{3\sigma_0}{256\sqrt{\pi}}\sum e_f^2\,\sqrt{Q^2\,Q_{\mathrm{sat}}^2\,(x)} \,,  & \mbox{for $\gamma =1/2$ (BFKL) }\,, 
\end{array} \right.
\label{FLlargeQ2}
\end{eqnarray} 

The analytical results above give the theoretical expectation for the large $Q^2$ region of the recent extraction of $F_L$ at fixed energy $W_{\gamma p}=276$ GeV. As $x\simeq Q^2/W_{\gamma p}^2$, the saturation scale is running with virtuality and hence one has $F_L(W_{\gamma p},Q^2) \propto (Q^2)^{-\lambda}\approx Q^{-0.6}$ for approaches having $\gamma=1$ at large virtualities. For the approaches based on LO BFKL evolution equation, one obtains $F_L (W_{\gamma p},Q^2)\propto (Q^2)^{(1-\lambda)/2}\approx Q^{\,0.7}$. This mild growth on virtuality is ruled out by data and the reason is that BFKL approach for structure functions should be valid just at intermediate virtualities. In fact, it is the transition for the double logarithmic evolution at small-$x$ and large $Q^2$. For a pure scaling fit with $\gamma = 0.84$ done in Ref. \cite{}, the behavior is still consistent and would give $F_L (W_{\gamma p},Q^2)\propto (Q^2)^{1-0.84(1+\lambda)}\approx Q^{-0.2}$. Therefore, the slope of the longitudinal structure function at fixed energy and large $Q^2$ could help to theoretically single out the correct effective anomalous dimension.

Now, let us  look at the turn over region. This limit corresponds to the intermediate and low $Q^2\leq 10$ GeV$^2$. This kinematical region is known to be in the geometric scaling region and the average dipole sizes are close to the saturation radius $R_{\mathrm{sat}}(x)$. The effective anomalous dimension is then given by the BFKL anomalous dimension at the saturation border $\gamma_{\mathrm{sat}}=0.63$. Therefore, at fixed $x$ the longitudinal structure function behaves as $F_L \propto (Q^2)^{1-\gamma_{\mathrm{sat}}}(Q_{\mathrm{sat}})^{\gamma_{\mathrm{sat}}}$. On the other hand, at fixed energy once again the saturation scale is running on $Q^2$ and thus one has $F_L (W_{\gamma p},Q^2)\propto (Q^2)^{1-0.63(1+\lambda)}\approx Q^{\,0.2}$. This corresponds to a mild growth on $Q^2$ near to the turn over point. We can also try to estimate where the turn over point should be located. In order to do so, let us expand the dipole cross section around the value $\rr^2\,Q_{\mathrm{sat}}^2/4=1$. For the expansion, we take the functional form of GBW for sake of simplicity and consider just the two first term in its Taylor series, $\sigma_{dip}\approx \sigma_0(1-2/e)+\sigma_0\,(\rr^2\,Q_{\mathrm{sat}}^2/4\,e)$. These two terms are the main contribution to $F_L$ in that region and the result can be obtained using Eq. (\ref{analytic}). The next term would give the twist-4 contribution, which is a negative correction. For fixed $x$, the longitudinal structure function will have the form,
\begin{eqnarray}
F_L\,(x,\,Q^2) \approx \frac{Q^2}{4\pi^2\alpha_{\mathrm{em}}}\,\left[\frac{\alpha_{\mathrm{em}}\bar{e}^2}{\pi}\sigma_0\left(1-\frac{2}{e}\right) + \frac{\alpha_{\mathrm{em}}\bar{e}^2}{\pi}\frac{\sigma_0}{e}\left(\frac{Q_{\mathrm{sat}}^2}{Q^2}\right)\right]\,,
\end{eqnarray}
where $\bar{e}^2 = \sum e_f^2$. Therefore, the turn over region can be obtained by finding the point of maximum for the function above. It can be shown that it is located at $Q^2 \propto Q_{\mathrm{sat}}^2(x)$, with the proportionality constant being dependent on the prefactors of the dipole cross section expansion. 

Finally, it is timely to investigate the effective anomalous dimension as a function of $Q^2$, taking the association $\rr^2\approx 4/Q^2$ for sufficiently large dipoles. It can be extracted from theoretical models by computing the following quantity, $\gamma_{\mathrm{eff}} = \frac{d\ln {\cal N}(rQ_{\mathrm{sat}},Y)}{d\ln(r^2Q_{\mathrm{sat}}^2/4)}$ where the dipole scattering amplitude is related to the dipole cross section by ${\cal N}(x,r) =\sigma_{dip}(x,r)/\sigma_0$. The analytical results for GBW, IIM and KKT parameterizations read as,
\begin{eqnarray}
\gamma_{\mathrm{eff}}^{\mathrm{GBW}}\,(\rr Q_{\mathrm{sat}},\,Y) & = & \frac{(\rr^2Q_{\mathrm{sat}}^2)\exp \left(-\frac{\rr^2Q_{\mathrm{sat}}^2}{4} \right)}{4\left[1-\exp \left(-\frac{\rr^2Q_{\mathrm{sat}}^2}{4}\right) \right]}\,,\\
\gamma_{\mathrm{eff}}^{\mathrm{IIM}}\,(\rr Q_{\mathrm{sat}},\,Y) 
& = & \left\{ \begin{array}{ll} 
\gamma_{\mathrm{sat}} - \frac{\ln \,(\rr^2Q_{\mathrm{sat}}^2/4)}{\kappa \,\lambda \,Y} \,,& \mbox{for $\rr Q_{\mathrm{sat}}\geq 2$}\,,\\
 a \ln (b\rr Q_{\mathrm{sat}})\exp \,[-a\ln^2\,(b\,\rr\, Q_{\mathrm{sat}})]\,,  & \mbox{for $\rr Q_{\mathrm{sat}}< 2$}\,.
\end{array} \right.\\
\gamma_{\mathrm{eff}}^{\mathrm{KKT}}\,(\rr Q_{\mathrm{sat}},\,Y) & = & \frac{C_F}{N_c}\,\bar{\gamma}\,\frac{(\rr^2Q_{\mathrm{sat}}^2)^{\bar{\gamma}}\exp \left[-\frac{C_F}{N_c}\left(\frac{\rr^2Q_{\mathrm{sat}}^2}{4}\right)^{\bar{\gamma}} \right]}{4^{\gamma}\left\{1-\exp \left[-\frac{C_F}{N_c}\left(\frac{\rr^2Q_{\mathrm{sat}}^2}{4}\right)^{\bar{\gamma}}\right] \right\}}\,,\hspace{0.5cm}\bar{\gamma}=\gamma\,(x,\,Q^2)\,.
\end{eqnarray}

\begin{figure}[t]
\begin{tabular}{cc}
\epsfig{file=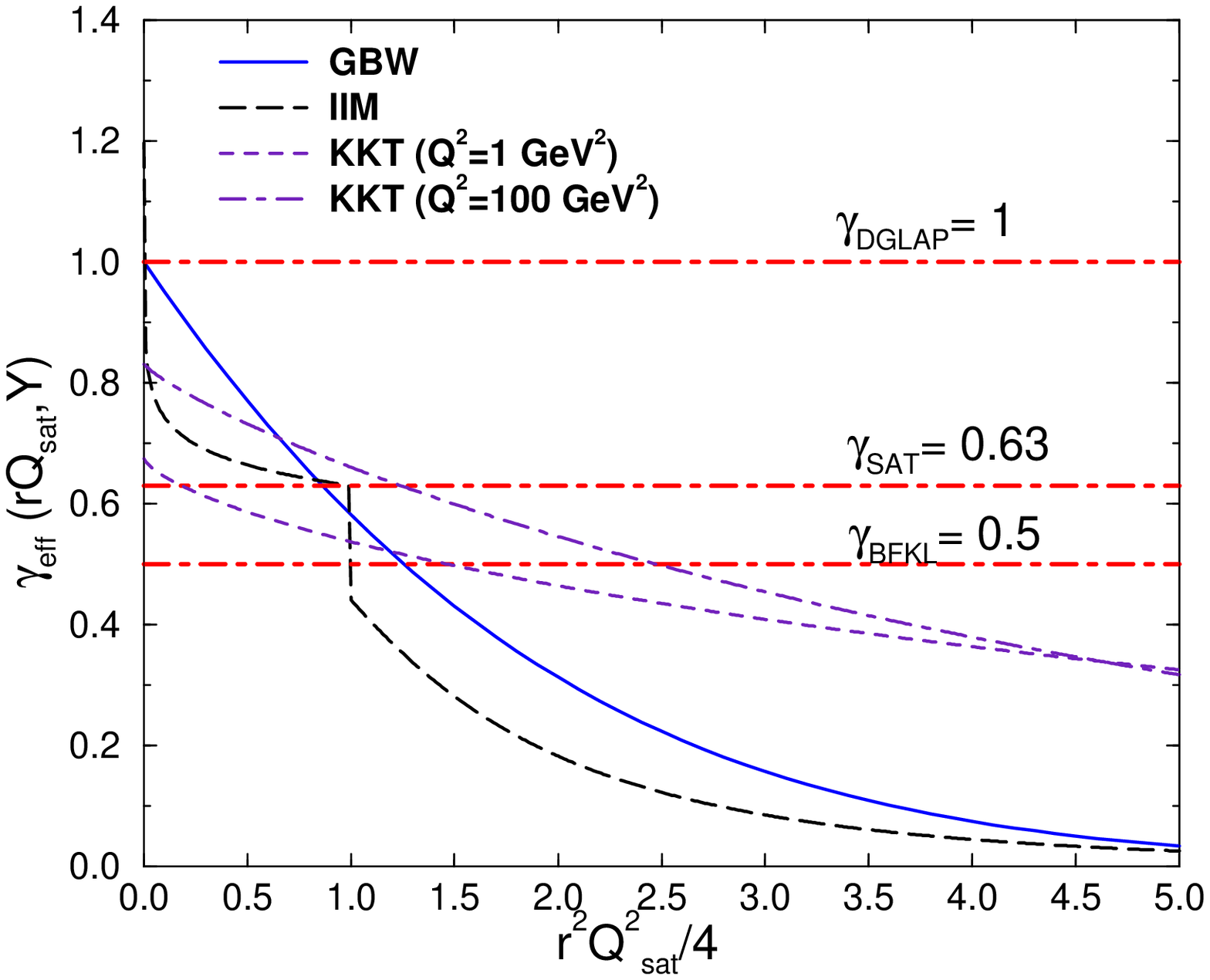,width=80mm} & \epsfig{file=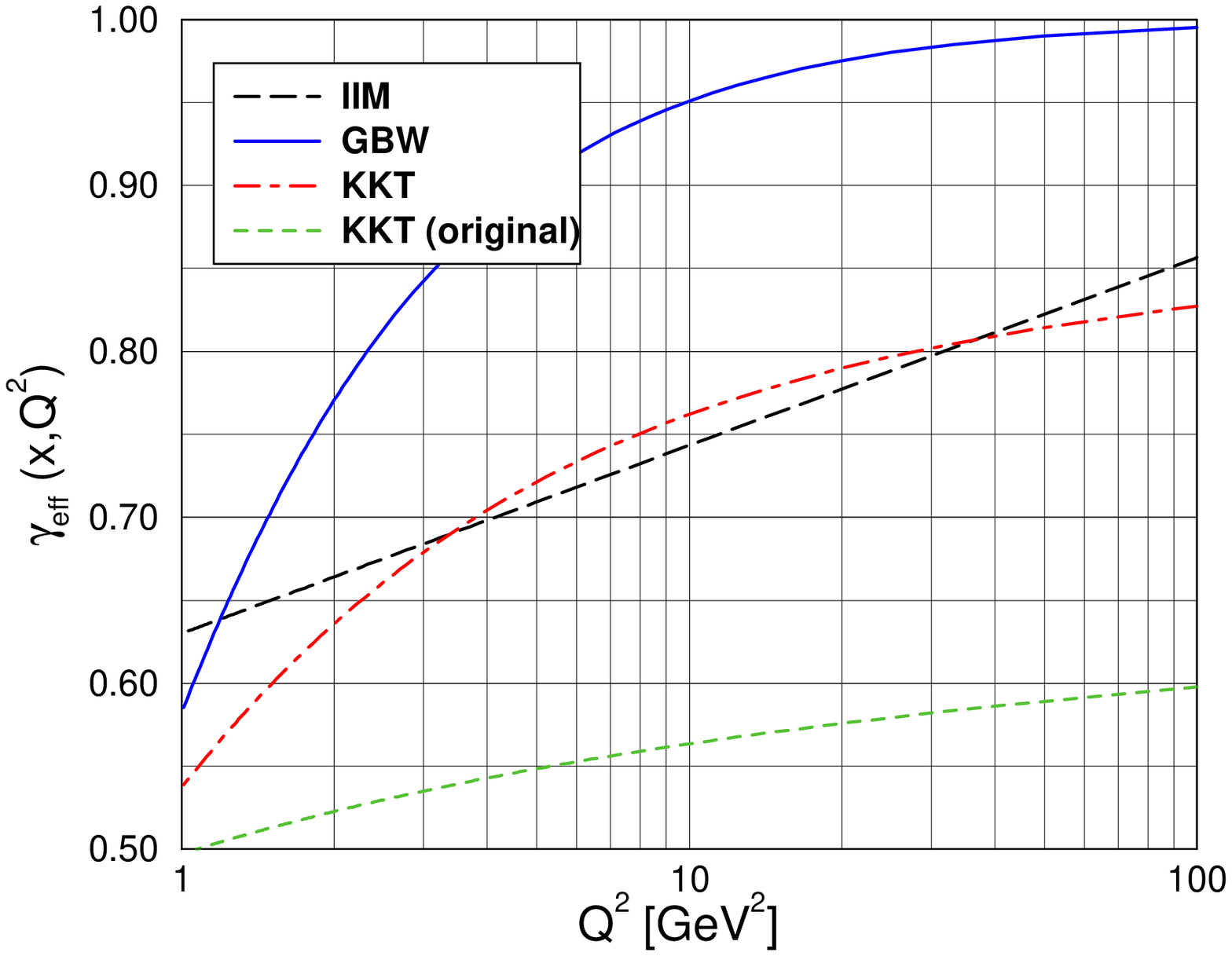,width=80mm}\\
(a) & (b)
\end{tabular}
\caption{\it (a) The effective anomalous dimension as a function of the quantity $\rr^2Q_{\mathrm{sat}}^2/4$ for different parameterizations. (b) The effective anomalous dimension as a function of $Q^2$.}
\label{figgamma}
\end{figure}

 In Fig. \ref{figgamma}-a, the effective anomalous dimension is plotted as a function of $\rr^2Q_{\mathrm{sat}}^2/4$. It should be noticed that the saturation region is placed at $\rr^2 Q_{\mathrm{sat}}^2/4>1$. The GBW model gives a smooth transition from $\gamma \rightarrow 0$ at the saturation region to $\gamma \rightarrow 1$ at dipole sizes $\rr \rightarrow 0$. The IIM model has a lower anomalous dimension than GBW in the saturation region, whereas they are similar at the scaling region $\rr Q_{\mathrm{sat}}\approx 2$. The transition to $\gamma =1$ is slowest than GBW and a larger region is driven by $\gamma_{\mathrm{sat}} = 0.63$ ($1 < \rr Q_{\mathrm{sat}} <2$). The discontinuity presented in the IIM plot is due to the step function for the interpolation between the different regions (saturation and color transparency). The behavior becomes steep in the limit $\rr \rightarrow 0$, strongly enhanced by the logarithmic dependence. The IIM parameterization violates scaling and for the plot we have considered $x=3\cdot 10^{-4}\,(Y=8)$, which corresponds to saturation scaling being the unit. In fact, the effective anomalous dimension decreases with $Y$ but at the same time is enhanced by the logarithmic contribution. The KKT parameterization presents also a slow transition to $\gamma \rightarrow 1$, which depends on the virtuality through the identification $\rr \approx 1/Q^2$ in the expression for $\bar{\gamma}$. In order to show this dependence, we have computed the effective anomalous dimension for the typical values $Q^2=1$ GeV$^2$ and 100 GeV$^2$ and fixed $x=3\cdot 10^{-4}$. The obtained values are close to the saturation and BFKL anomalous dimension for a large span of dipole sizes. For $Q^2=1$ GeV$^2$, in the small dipole limit the KKT effective anomalous dimension goes to 0.68 whereas for 100 GeV$^2$ it goes to 0.84. In Fig. \ref{figgamma}-a, for completeness we also show the DGLAP, saturation and BFKL values for the anomalous dimension.

In Fig. \ref{figgamma}-b is shown the effective anomalous dimension as a function of the virtuality $Q^2$, using the average dipole size as $\rr=2/Q$. The GBW parameterization presents a fast convergence for the DGLAP anomalous dimension at large $Q^2$. On the other hand, both IIM and KKT parameterizations have a mild growth on virtuality and they converge for a value $\gamma \approx 0.85$ at large $Q^2$. At intermediate virtualities, the GBW results have sizeable deviations in relation to IIM and KKT. At lower virtualities $Q^2=1$ - 2 GeV$^2$, the anomalous dimension reaches 0.63 for IIM and 1/2 for KKT. For completeness, it is shown also the KKT result using its original parameters. The effective anomalous dimension is very low in a large span of $Q^2$, having an average value near the BFKL anomalous dimension. This fact explains the known impossibility of describing intermediate and large $Q^2$ DESY-HERA data with the original KKT model. Its success in describing RHIC data on charged hadron spectra for $d-Au$ is due to the low $p_T\leq 5$ GeV range measured there and large mid-rapidity $y \approx 0$, where the saturation scale is not too large. It should be noticed that for the  original KKT parameterization the saturation scale is extremely large, reaching about $Q_{\mathrm{sat}}^2\approx 4$ at $x=10^{-4}$. 

In short, the different anomalous dimension for each parameterization of the dipole cross section should account for a distinct $Q^2$ behavior for the longitudinal structure function at fixed energy. Namely, the turn over point and the large $Q^2$ region can provide valuable information. This is studied in details in next section.

\section{Results and Discussions}
\begin{figure}[t]
\centerline{\psfig{file=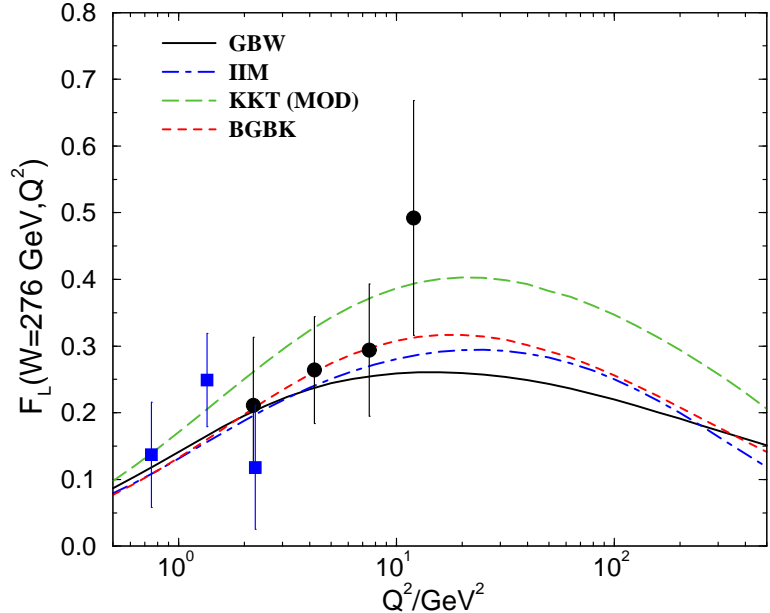,width=100mm}}
 \caption{\it The results for $F_L(x,Q^2)$ as a function of $Q^2$ at fixed energy $W_{\gamma p} = 276$ GeV. The different numerical results correspond to distinct dipole cross section parameterizations (see text).}
\label{fig1}
\end{figure}

Lets present the numerical results for the distinct parameterization at fixed energy $W_{\gamma p}=276$ GeV as a function of $Q^2$, which are shown in  Fig. \ref{fig1}. They are compared to the preliminary experimental data \cite{h1_prel}, namely the 99 min. bias data (circles) and 2000 shifted vertex data (squares). The total error for $F_L$ is shown. The models give a reasonable data descriprion. The GBW parameterization gives the lower result and having the smaller turn over point, around $Q^2 = 20$ GeV$^2$. It presents also a less steep behavior at large virtualities. It has been shown it successfully describes the preliminary H1 data in Ref. \cite{h1_prel}. Therefore, a direct comparison among the several models is timely and necessary. The main shortcoming in the experimental measurements is large uncertainties leading to considerable systematic/statistic errors. More precise data and higher statistics would help to distinguish between different models. The IIM parameterization has a turn over point placed at at larger $Q^2=30$ GeV$^2$ and has a steeper fall at large $Q^2$. Similar feature is present in KKT and BGBK models. This is due to the similar anomalous dimension for them in that region. The same particularity is present at the low $Q^2$, where all parameterization have anomalous dimension near $\gamma_{\mathrm{sat}}$. The modification in KKT parameters presented in the previous section leads to numerical results having similarities with to BGBK up to a different overall normalization. This can be contrasted with $F_2$ data and its universality could be tested. 

\begin{figure}[t]
\centerline{\psfig{file=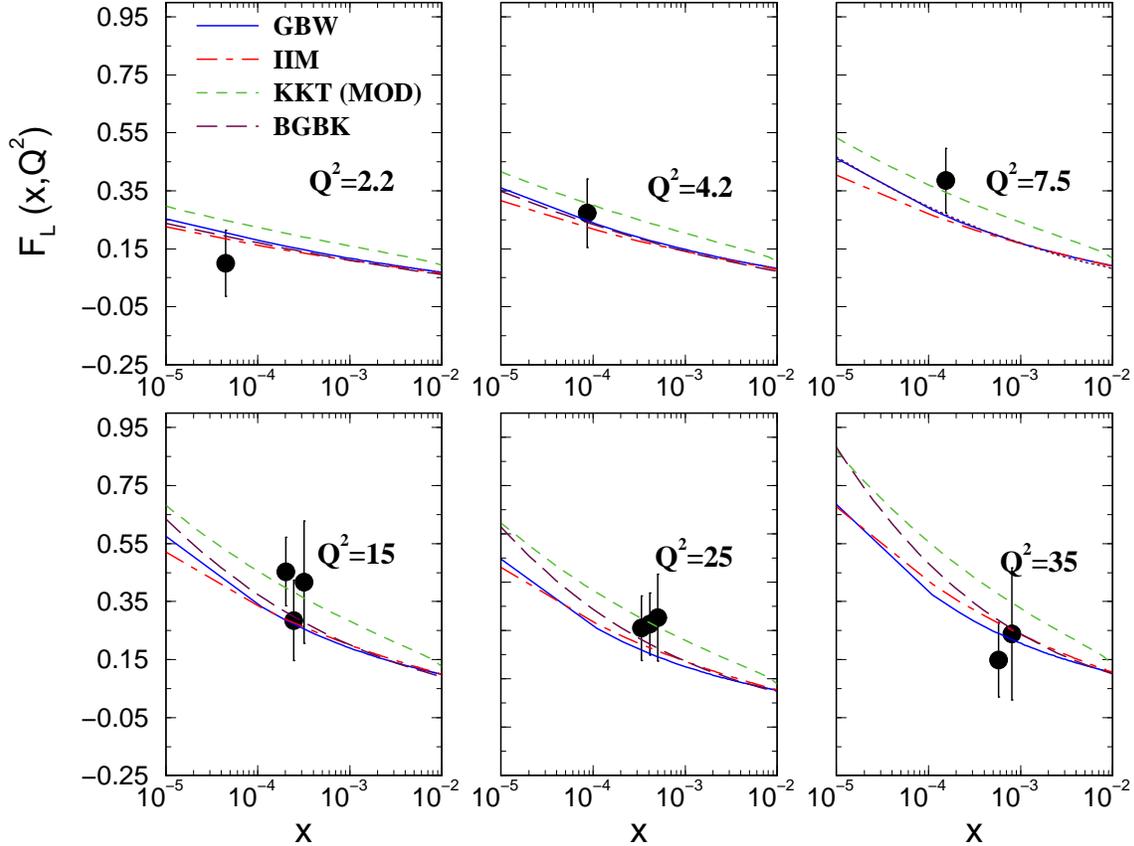,width=150mm}}
 \caption{\it The results for $F_L(x,Q^2)$ as a function of $x$ at fixed $Q^2$. Data from H1 Collaboration.}
\label{fig_comp_dipoles}
\end{figure}

\begin{figure}[t]
\centerline{\psfig{file=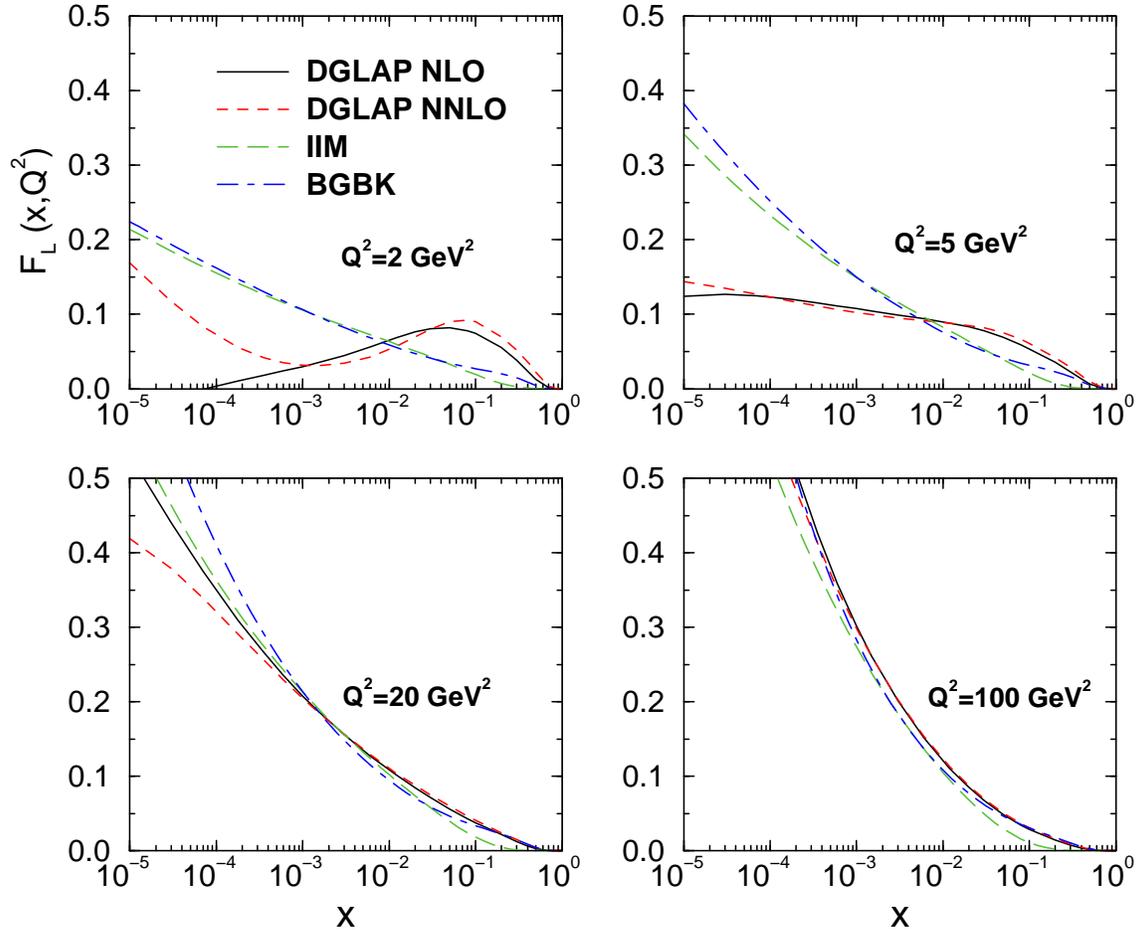,width=150mm}}
 \caption{\it Comparison among color dipole results and recent NLO/NNLO DGLAP analysis.}
\label{comp_dglap}
\end{figure}

It is also timely to plot $F_L$ as a function of $x$ at fixed values of $Q^2$, as this is the usual way it is presented. It would be useful in order to figure out what are the differences in the $x$-behavior for different dipole models considered here. In Fig. \ref{fig_comp_dipoles}, we show the results for the distinct dipole models in comparison with the experimental data \cite{h1fldata}. The IIM and GBW models produce similar results for all $Q^2$. Deviations are larger for BGBK at intermediate $Q^2$ and very small $x$. The modified KKT model (with ad hoc parameters discussed in previous section) produces a consistent description, mostly at low $Q^2$. It can be improved by using a smaller overall normalization, which it would bring the theoretical curves close to experimental results.
It is important compare these results from the color dipole approach against the more recent DGLAP QCD analysis. For doing so, we consider the last analysis on $F_L$ using NLO and NNLO accuracy \cite{MST_nnlo}, which include the reduced cross section at high $y$ and its earlier direct measurements. For comparison, we have chosen the IIM and BGBK dipole models. The reason is IIM being the more recent fit using saturation physics and it interpolates BFKL physics at large $Q^2$. The BGBK model contains DGLAP evolution at large $Q^2$ and saturation corrections at small-$x$ and low $Q^2$. For IIM, we include a multiplicative threshold factor $(1-x)^5$ to correctly describe the limit $x\rightarrow 1$. The results are shown in Fig. \ref{comp_dglap}, for the sample virtualities $Q^2=2,\,5,\,20,\,100$ GeV$^2$. There are strong deviations for low $Q^2$, whereas the results are close at larger virtualities. At that region, the approaches are all convergent. The dipole approach gives numerical results which are between the LO and NLO/NNLO DGLAP analysis and are consistent with the independent dipole model type fit in Ref. \cite{Thorne_dipole}. Concerning higher twist contributions, in Ref. \cite{MST_nnlo} it is implemented via renormalon correction in the non-singlet sector. However, they basically remain the same behavior on $x$ as the NLO/NNLO analysis. The deviation is larger at  low-$Q^2$, differing just by a distinct overall normalization. The dipole results are different from the renormalon correction as color dipoles models contain both higher-order corrections and higher twists.

\begin{figure}[t]
\centerline{\psfig{file=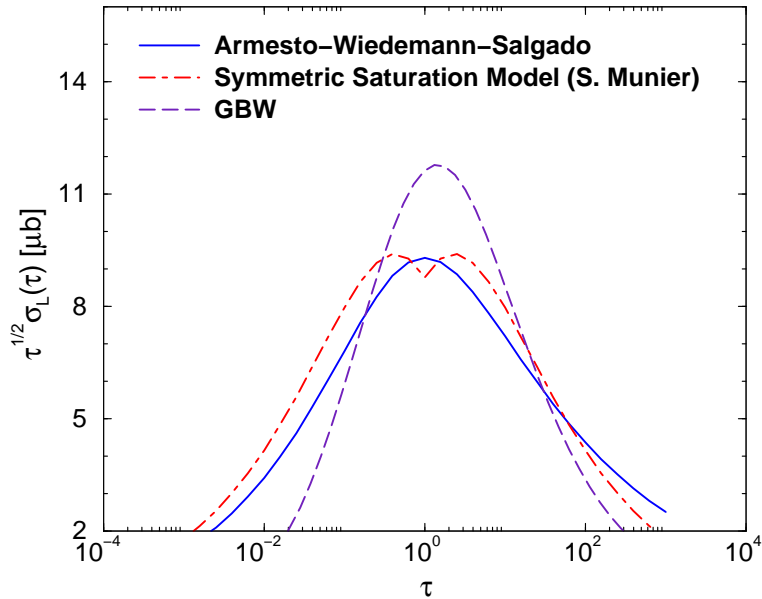,width=100mm}}
 \caption{\it The geometric scaling pattern for the longitudinal $\gamma^*p$ cross section. The theoretical predictions are taken from scaling functions for the total inclusive cross section (see text).}
\label{fig2}
\end{figure}

An interesting question concerning the longitudinal cross section is whether it exhibits geometric scaling property as the total inclusive cross section. Namely, the DIS cross section at high energies depends on $x$ and $Q^2$ exclusively via the scaling variable $\tau_p = Q^ 2/Q_{\mathrm{sat}}^2$, i.e. $\sigma_{tot}^{\gamma^*p}(x,\,Q^2)=\sigma_{tot}^{\gamma^*p}\,(\tau_p)$. Within the color dipole formalism and parton saturation approaches this property should be present at small-$x$ data. In order to address this issue, we consider two simple parameterizations for geometric scaling in the inclusive cross section and extrapolate the results for the longitudinal contribution. At low $x$ and within the color dipole picture, the ratio between the longitudinal and total structure function can be roughly approximated by $F_L/F_2\approx 2/11$ \cite{magsasha}. For the functional form of the scaling function in DIS we take the results from Armesto-Wiedemann-Salgado in Ref. \cite{Armesto_scal}, where the high energy lepton-hadron, proton-nucleus and nucleus-nucleus collisions have been related through geometric scaling. The following scaling curve for the photoabsortion cross section has been considered based on theoretical motivations:
\begin{eqnarray}
  \sigma^{\gamma^* p}_{tot}\,(\tau_p) = \bar{\sigma}_0\,
  \left[ \gamma_E + \Gamma\left(0,\beta \right) +
         \ln \left(\beta\right) \right],\hspace{0.5cm} \beta = a/\tau_p^{b}\,,
       \label{sigtot_param_tau}
\end{eqnarray}
where $\gamma_E$ is the Euler constant and $\Gamma\left(0,\beta\right)$
the incomplete Gamma function. The parameters for the proton  case were obtained from a fit to the small-$x$ $ep$ DESY-HERA data, producing $a=1.868$, $b=0.746$ and the overall  normalization was fixed by $\bar\sigma_0=40.56$ $\mu$b. The saturation scale is taken from the GBW parameterization. In Fig. \ref{fig2} we plot $\sqrt{\tau_p}\,\sigma_L\,(\tau_p)$ as a function of $\tau_p$, where we have used  the approximation $\sigma_L\approx (2/11)\,\sigma_{tot}^{\gamma^*p}$. The functional behavior remains the same as for the total inclusive case. Experimentalists could study this property using the available data at small-$x$ for $F_L$ and look for a geometric scaling pattern.

The  symmetric form of the scaling cross section has been also addressed in Ref. \cite{Muniersymm}, where the symmetry between low and high $Q^2$ region is related to the symmetry of the two-gluon-exchange dipole-dipole cross section. It has been assumed that in the dipole frame the energy evolution leads to the multiplication of partons and consequently to the appearance of the saturation scale $Q_{\mathrm{sat}}^2=\Lambda^2 e^{\lambda\log(1/x)}$. When the virtuality is larger than the saturation scale, one has the usual DIS picture with the photon probing a set of independent partons. The growth rate of the parton densities is given by $Q_{\mathrm{sat}}^2/\Lambda^2$, where $\Lambda$ is such  that the area of the proton is of order $1/\Lambda^2$ throughout the evolution. On the other hand, when the virtuality is close to saturation scale, the probability of multiple interactions between the photon and the proton becomes sizeable and a Glauber-like resummation is introduced and extrapolated by symmetry for virtualities lower than the saturation scale. Explicitly, the functional form for the scaling curve in the symmetric saturation model reads as,
\begin{eqnarray}
\sigma_{tot}^{\gamma^*p}\,(\tau_p)= \left\{ \begin{array}{ll} 
\frac{N}{\Lambda^2}\,\frac{1}{a}\left\{1-\exp \,\left[-\frac{a}{\tau_p}\left(1+ \log \sqrt{\tau_p}\right)  \right] \right\} \,,& \mbox{for $Q^2\geq Q_{\mathrm{sat}}$}\,,\\
 \frac{N}{\Lambda^2}\,\frac{1}{b\,\tau_p}\left\{1-\exp \,\left[-\,b\,\tau_p\left(1-\log \sqrt{\tau_p}\right)  \right] \right\}\,,  & \mbox{for $Q^2 <Q_{\mathrm{sat}}$}\,.
\end{array} \right.\,,
\end{eqnarray}
where $N$ (the normalization factor), $a$ and $b$ are free parameters fixed by DIS data. The result is plotted in Fig. \ref{fig2} for the longitudinal case.  It is similar to previous parameterization, with deviations for large $Q^2$. This is due the limit of both functional forms at that limit. For completeness, we calculate the complete prediction from the GBW parameterization in order to check the approximation $F_L/F_2=2/11$ considered here. The approximation is reasonable in the intermediate $Q^2$ region whereas it overestimates the ratio at low $Q^2$ by about 30\%.

As a summary, we have analyzed the longitudinal structure function at fixed energy within the dipole formalism including parton saturation effects. These effects are important because they resume  higher twist corrections to the process. In particular, we have shown the recent experimental data could help on investigating the effective anomalous dimension. The regions of interest are the large $Q^2$ and the turn over point. We have computed expressions for the anomalous dimension for distinct analytical parameterizations for the dipole cross section and show its role in the $Q^2$ behavior. A precise measurement in the turn over point would allow distinguish among different models. It is shown that the longitudinal cross section should exhibit geometric scaling property as a natural extension as for the inclusive total cross section. This can be studied by experimentalists using the available measurements. Therefore, $F_L$ is an outstanding observable testing both parton saturation and twist resummation and more precise data and/or more statistics is increasingly desirable.

\section*{Acknowledgments}
The author is indebted to R.S. Thorne for his help on the recent results on NLO/NNLO DGLAP analysis of $F_L$. The author thanks N. Nikolaev and M. Lublinsky for their comments and for calling attention for valuable references.  N.P. Zotov and E.~M.~Lobodzinska are gratefully acknowledged for discussions on $k_{\perp}$-factorization results and  information on recent H1 Coll. data, respectively. This work has received  support of the High Energy Physics
Phenomenology Group, GFPAE IF-UFRGS (Brazil).

\end{document}